\documentclass[10pt,journal,twoside,onecolumn,romanappendices]{IEEEtran}

\usepackage{psfrag,epsfig,graphics}
\usepackage{amsmath,amssymb,amsthm,multirow}
\usepackage{cite,subfigure,balance}

\usepackage[ruled, linesnumbered, vlined]{algorithm2e}
\usepackage{bm}

\usepackage{color}

\definecolor{violet}{rgb}{0.5,0,0.5}

\newcommand{\cb}[1]{{\boldsymbol{#1}}}
\newcommand{\cp}[1]{\ifmmode {\mathcal{#1}}\else ${\mathcal{#1}}$\fi}

\newcommand{\bbeta}{\boldsymbol{\beta}}

\newcommand{\bH}{\boldsymbol{H}}
\newcommand{\bI}{\boldsymbol{I}}

\newcommand{\bQ}{\boldsymbol{Q}}
\newcommand{\bR}{\boldsymbol{R}}

\newcommand{\bu}{\boldsymbol{u}}

\newcommand{\bp}{\boldsymbol{p}}

\newcommand{\bs}{\boldsymbol{s}}

\newcommand{\bw}{\boldsymbol{w}}
\newcommand{\btw}{\tilde{\boldsymbol{w}}}

\newcommand{\bx}{\boldsymbol{x}}

\newcommand{\tr}{\text{tr}}

\newcommand{\E}{{\mathbb{E}}}

\usepackage{tikz}

\begin{document}

\title{Adaptive parameters adjustment \\ for group reweighted zero-attracting LMS}

\author{Danqi Jin,\; Jie Chen,\; C{\'e}dric Richard,\; Jingdong Chen

\thanks{The work of Jie Chen was supported in part by NSFC grant 61671382. The work of C{\'e}dric Richard was supported in part by ANR and in part by DGA under Grant ANR-13-ASTR-0030 (ODISSEE Project). The work of Jingdong Chen was supported in part by NSFC grant 61425005.}
\thanks{D. Jin, J. Chen and J. Chen are with Centre of Intelligent Acoustics and Immersive Communications at School of Marine Science and Technology, Northwestern Polytechinical University, Xi'an, China (emails: danqijin@mail.nwpu.edu.cn, dr.jie.chen@ieee.org, jingdongchen@ieee.org). C. Richard is with Laboratoire Lagrange, Universit\'e C\^ote d'Azur, CNRS, OCA, France (email: cedric.richard@unice.fr). }}

\maketitle

\begin{abstract}
Group zero-attracting LMS and its reweighted form have been proposed for addressing system identification problems with structural group sparsity in the parameters to estimate. Both algorithms however suffer from a trade-off between sparsity degree and estimation bias and, in addition, between convergence speed and steady-state performance like most adaptive filtering algorithms. It is therefore necessary to properly set their step size and  regularization parameter. Based on a model of their transient behavior, we introduce a variable-parameter variant of both algorithms to address this issue. By minimizing their mean-square deviation at each time instant, we obtain closed-form expressions of the optimal step size and regularization parameter. Simulation results illustrate the effectiveness of the proposed algorithms.
\end{abstract}


\section{Introduction}
\label{sec:intro}

Adaptive filtering algorithms are powerful tools for online system identification~\cite{Sayed2008,Widrow1985}. Among the set of existing solutions, the least-mean-squares (LMS) algorithm plays a central role due to its robustness, good performance and low complexity. A number of variants have been proposed in the literature to endow the LMS with useful additional characteristics such as the ability to promote sparse estimates. This property is required in applications such as channel identification where, though the impulse responses can be long, only a few coefficients may have significant values. Several algorithms were derived, such as the proportionate NLMS (PNLMS)~\cite{Duttweiler2000,Benesty2002}, the zero-attracting LMS algorithm (ZA-LMS) and its reweighted form~(RZA-LMS)~\cite{Chen2009ZALMS}. They offer improved performance in sparse scenarios provided that their parameters are appropriately set.

Beyond element-wise sparsity, a further consideration is that some sparse systems may be group-sparse~\cite{Chen2012Recursive,Chen2010RegularizedLMS}. Compared to general sparse systems whose impulse response is not necessarily structured, group-sparse systems have an impulse response composed of a few distinct clusters of nonzero coefficients. Applications are abundant, e.g., specular multipath acoustic and wireless channels estimation~\cite{Chen2012Recursive,Chen2010RegularizedLMS,Schreiber1995}. Using such structural prior information should lead to enhanced performance. Based on mixed norm regularization, the $\ell_{1,\infty}$-regularized RLS~\cite{Chen2012Recursive}, the group ZA-LMS (GZA-LMS) and the group reweighted ZA-LMS (GRZA-LMS)~\cite{Chen2010RegularizedLMS} were proposed to promote group-sparsity. Nevertheless, setting the algorithm parameters such as the step size and the regularization parameter to ensure a performance gain remains a tricky task~\cite{Chen2012Recursive,Chen2010RegularizedLMS,Huang2010}. On the one hand, the step size plays a critical role in the trade-off between the convergence speed and the steady-state performance. On the other hand, the regularization parameter controls the trade-off between the degree of sparsity and the estimation bias. It is worth noting that setting one of these parameters to an inappropriate value may deteriorate the estimation performance.

Variable parameter strategies usually provide an efficient way to achieve a reasonable trade-off between competing performance requirements~\cite{Paleologu2008}. Several variable step size strategies have been proposed for the LMS and ZA-LMS algorithms~\cite{Paleologu2008,Salman2012,Fan2017,Benesty2006,Kwong1992}, mostly based on estimation error. There are however few works addressing this issue for group-sparse LMS. Motivated by our recent work~\cite{Chen2016ZALMS}, we propose in this paper to derive the so-called variable-parameter GZA-LMS (VP-GZA-LMS) and GRZA-LMS (VP-GRZA-LMS) algorithms. The step size and the regularization parameter are both adjusted in an online way, based on an optimization procedure that minimizes the mean-square deviation (MSD) at each iteration. Experiments illustrate the effectiveness of this strategy, which leads to a faster convergence rate as well as a lower misadjustement error.

\textbf{Notation.} Normal font $x$ denotes scalars.  Boldface lowercase letters $\bx$ and uppercase letters $\cb{X}$ denote column vectors and matrices, respectively. The superscript ${(\cdot)}^\top$ and ${(\cdot)}^{-1}$ denote the transpose and inverse operators, respectively. $\cb{0}_N$ and $\cb{1}_N$ denote all-zero vector and all-one vector of length $N$. The operator $\text{tr}\{\cdot\}$ takes the trace of its matrix argument. The mathematical expectation is denoted by $\E\{\cdot\}$. The Gaussian distribution with mean $\mu$ and variance $\sigma^2$ is denoted by $\cp{N}(\mu,\sigma^2)$. The operator $\max\{\cdot, \cdot\}$ and $\min\{\cdot, \cdot\}$ take the maximum or minimum of their arguments.

\section{SYSTEM MODEL AND GROUP-SPARSE LMS}
\label{sec:format}

Consider an unknown system defined by the linear model:
\begin{equation}
	\label{eq:linear.model}
	d_n = \bu_n^\top\bw^{\star} + z_n
\end{equation}
at time instant $n$, where $\bw^{\star} \in \mathbb{R}^L$ is an unknown parameter vector, $\bu_n \in \mathbb{R}^L$ is a zero-mean regression vector with positive definite covariance matrix, and $d_n$ is the output signal assumed to be zero-mean. The error signal $z_n$ is assumed to be stationary, independent and identically distributed (i.i.d.), with zero-mean and variance~$\sigma_z^2$, and independent of any other signal.

Consider the mean-square error (MSE) cost $J(\bw)$, namely,
\begin{equation}
	\label{eq:MSE}
	J(\bw) =\frac{1}{2}\,\mathbb{E}\big\{[d_n - \bw^\top\bu_n]^2 \big\}
\end{equation}
It can be checked that $\bw^{\star}$ is the minimizer of $J(\bw)$. In this paper we consider the problem of estimating the unknown parameter vector~$\bw^\star$ when it is group-sparse. This problem can be addressed by minimizing the following regularized MSE cost:
\begin{equation}
	\label{eq:mse.za}
	\begin{split}
	\bw^{\rm o}_{\rm GZA} &= \arg\min_{\bw} J_{\rm GZA} (\bw) \\
	\text{with }\, J_{\rm GZA} (\bw)  &= \frac{1}{2}\,
    \mathbb{E}\big\{[d_n - \bw^\top\bu_n]^2 \big\} + \lambda \|\bw\|_{1,2}
	\end{split}
\end{equation}
where $\lambda\geq 0$ is the regularization parameter. The $\ell_{1,2}$-norm of $\bw$, defined below, allows to promote its group-sparsity:
\begin{equation}
	\|\bw\|_{1,2} = \sum_{j=1}^J{\|\bw_{{\cal G}_j}\|}_2
\end{equation}
where ${\{{\cal G}_j\}}_{j=1}^J$ is a partition of the index set ${\cal G} = \{1, \ldots, L\}$, and $\bw_{{\cal G}_j}$ denotes the subvector of $\bw$ with entries indexed by ${\cal G}_j$. Calculating a subgradient of $J_{\rm GZA} (\bw)$, then approximating second-order moments by instantaneous estimates, leads to the following subgradient update equation in subvector form:
\begin{equation}
           \label{eq:update.GZALMS}
            \bw_{n+1, {\cal G}_j} = \bw_{n, {\cal G}_j} + \mu \, e_n  \bu_{n, {\cal G}_j} - \rho \,\, \bs_{n, {\cal G}_j}
\end{equation}
for $j\in\{1,\ldots,J\}$, with:
\begin{equation}
    \label{eq:s}
    \bs_{n, {\cal G}_j} =
        \begin{cases}
        \frac{\bw_{n,{\cal G}_j}}{{\|\bw_{n,{\cal G}_j}\|}_2}\quad
        &\text{for ${\|\bw_{n,{\cal G}_j}\|}_2\neq0$}\\
        \quad\,\,0\quad&\text{for ${\|\bw_{n,{\cal G}_j}\|}_2 = 0$,}
        \end{cases}
\end{equation}
with $e_n = d_n - \bw_n^\top \bu_n$ the estimation error, $\bu_{n, {\cal G}_j}$ the subvector of $\bu_n$ with entries indexed by ${\cal G}_j$, $\mu$ a positive step size, and $\rho = \mu \lambda$ the shrinkage parameter.

The GRZA-LMS was proposed to reinforce group-sparsity and then get enhanced performance in group-sparse system identification. Consider the optimization problem:
\begin{equation}
	\label{eq:mse.rza}
	\begin{split}
	\bw^{\rm o}_{\rm GRZA} &= \arg\min_{\bw} J_{\rm GRZA} (\bw) \\
	\text{with }\, J_{\rm GRZA} (\bw)  &= \frac{1}{2}\,
    \mathbb{E}\big\{[d_n - \bw^\top\bu_n]^2 \big\}+\lambda\sum_{j=1}^J\log\Big[1+\frac{\|\bw_{{\cal G}_j}\|_2}
    {\varepsilon}\Big]
	\end{split}
\end{equation}
where the log-sum penalty term is used to make group-sparsity attractor takes effort only for groups at the same level as $\varepsilon$~\cite{Chen2009ZALMS}. Similarly, using a stochastic subgradient update yields the GRZA-LMS:
\begin{equation}
           \label{eq:update.GRZALMS}
            \bw_{n+1, {\cal G}_j} = \bw_{n, {\cal G}_j} + \mu \, e_n  \bu_{n, {\cal G}_j} - \rho \,\beta_{n,j} \bs_{n,{\cal G}_j},
\end{equation}
where $\beta_{n,j} = {1}/[{{\|\bw_{n,{\cal G}_j}\|}_2 + \varepsilon}]$ is a weighting coefficient. Equivalently, equation \eqref{eq:update.GRZALMS} in vector form is given by:
\begin{equation}
	\label{eq:update.GRZALMS.vector}
	\bw_{n+1} = \bw_{n} + \mu \, e_n  \bu_{n}
    - \rho \,\bbeta_{n}\circ\bs_{n},
\end{equation}
where $\bbeta_{n}$ and $\bs_{n}$ are vector forms of $\beta_{n,j}$ and $\bs_{n,{\cal G}_j}$, respectively, with dimension $L\times 1$. Symbol $\circ$ denotes the Hadamard product.

Observe from equations \eqref{eq:update.GZALMS} and \eqref{eq:update.GRZALMS} that  GRZA-LMS reduces to GZA-LMS by replacing the parameters $\beta_{n,j}$ with 1, that is,
\begin{equation}
    \label{eq:twocase}
    \beta_{n,j} =
        \begin{cases}
        \frac{1}{
        {\|\bw_{n,{\cal G}_j}\|}_2+\varepsilon}\quad&\text{GRZA-LMS}\\
        \quad\quad1\quad&\text{GZA-LMS.}
        \end{cases}
\end{equation}
We shall derive a variable parameter strategy for both GZA-LMS and GRZA-LMS algorithms based on the general form~\eqref{eq:update.GRZALMS.vector}, while specific algorithms can be obtained by setting $\beta_{n,j}$ according to~\eqref{eq:twocase}.

\section{MODEL-BASED PARAMETER DESIGN OF GRZA-LMS}
\label{sec:VPGRZALMS}
\subsection{Transient Behavior Model of GRZA-LMS}

Define the weight error vector $\tilde\bw_n$ as the difference between the estimated weight vector $\bw_n$ and $\bw^\star$:
\begin{equation}
	\label{eq:w.error}
	\tilde\bw_n = \bw_n - \bw^\star
\end{equation}
To derive our variable parameter strategy, we analyze the transient behavior of the mean-square deviation (MSD) of $\bw_n$ over time, defined as: $\xi_n=\mathbb{E}\{\|\tilde{\bw}_n\|^2\}$. To keep the calculations mathematically tractable, we introduce the independence assumption~\cite{Sayed2008}:

\textbf{A1}: The weight-error vector $\tilde\bw_n$ is statistically independent of the input vector $\bu_n$.

\noindent This assumption is commonly used in the adaptive filtering literature since it helps simplify the analysis, and the performance results obtained under this assumption match well the actual performance of filters for sufficiently small step sizes~\cite{Sayed2008}.

Subtracting $\bw^\star$ from both sides of \eqref{eq:update.GRZALMS.vector}, using $e_n = z_n - \tilde\bw_n^\top\bu_n$, leads to the update equation of $\tilde\bw_n$:
\begin{equation}
	\label{eq:w.error.upd.GRZALMS}
	\tilde\bw_{n+1} = \tilde\bw_{n} + \mu\bu_{n} z_n
    - \mu\bu_{n} \bu_n^\top \tilde\bw_n - \rho\bbeta_{n}\circ\bs_{n}.
\end{equation}

Using A1 and $e_n = z_n - \tilde\bw_n^\top\bu_n$, the MSE of the GRZA-LMS is given by:
\begin{equation}
	\label{eq:mse.ZALMS}
	\mathbb{E}\{e_n^2\} = \sigma_z^2 + \tr\{\bR_u\bQ_n\}
\end{equation}
with $\bQ_n = \mathbb{E}\{\tilde\bw_n\tilde\bw_n^\top\}$. The quantity $\tr\{\bR_u\bQ_n\}$ is the excess mean-square error (EMSE) at time instant $n$, denoted by $\zeta_n$. Note that $\xi_n = \tr\{\bQ_n\}$. Again, to keep the calculations mathematically tractable, we introduce the following assumptions~\cite{Sayed2008}:

\textbf{A2}: The input regressor $\bu_n$ is a zero-mean white signal with covariance matrix $\bR_u = \sigma_u^2 \bI$.

\textbf{A2'}: The input regressor $\bu_n$ is Gaussian  distributed.

Though introducing A2 and A2' simplify the derivation, as illustrated with simulation results, it turns out that the resulting algorithms work well in non-Gaussian correlated input scenarios where these assumptions do not hold. As shown in the sequel, A2' is only used with \eqref{eq:GRZA.a} to make the calculation of $g$ tractable. Under A2, we can relate the MSD to the EMSE via a scaling factor:
\begin{equation}
           \label{eq:MSD&EMSE}
            \zeta_n = \sigma_u^2\,\tr\{\bQ_n\}= \sigma_u^2\,\xi_n.
\end{equation}
We shall now determine a recursion for $\tr\{\bQ_n\}$ in order to relate the MSD at two consecutive time instants $n$ and $n+1$. Post-multiplying~ \eqref{eq:w.error.upd.GRZALMS} by its transpose, taking the expectation and matrix trace, using A1 and A2, we get:
\begin{equation}
           \label{eq:w.error.square.GRZALMS}
            \tr\{\bQ_{n + 1}\} = \tr\{\bQ_n\} + \mu^2g + \rho^2h + 2\mu\rho \ell - 2\mu r_1 - 2\rho r_2
\end{equation}
with
{\setlength\abovedisplayskip{-3pt}
\setlength\belowdisplayskip{-2pt}
\begin{align}
	g &= \sigma_z^2\ \tr\{\bR_u\} + \E\{\bu_n^\top\btw_n\btw_n^\top\bu_n\bu_n^\top\bu_n\}		\label{eq:GRZA.a}\\
	h &= \E\bigl\{(\bbeta_{n}\circ\bs_{n})^\top(\bbeta_{n}\circ\bs_{n})\bigr\}	\label{eq:GRZA.b}\\
	\ell & = \E\bigl\{\btw_n^\top\bu_n{\bu_{n}^\top(\bbeta_{n}\circ\bs_{n})}\bigr\}					\label{eq:GRZA.c}\\
	r_1 & = \E\bigl\{\btw_n^\top\bu_n\bu_n^\top\btw_n\bigr\}					\label{eq:GRZA.p1}\\
	r_2 & = \E\bigl\{{(\bbeta_n\circ\bs_n)}^\top\tilde\bw_{n}\bigr\}.	\label{eq:GRZA.p2}
 \end{align}}

\noindent We have dropped the time index $n$ in the left-hand side of \eqref{eq:GRZA.a}--\eqref{eq:GRZA.p2} for compactness.

\subsection{Parameter Design Using Transient Behavior Model}
\label{sec:DesignZA-LMS}

We shall now derive our parameter design strategy for GRZA-LMS using model \eqref{eq:w.error.square.GRZALMS}. Given the MSD $\xi_n$ at time instant $n$, we need to determine the parameters $\{\mu_n, \rho_n\}$  that minimize the MSD $\xi_{n+1}$:
\begin{equation}
	\label{eq:emse}
	\{\mu^\star_n, \rho^\star_n\} = \arg\min_{\mu, \rho} \, \xi_{n+1} \,|\, \xi_{n}.
\end{equation}
Using recursion~\eqref{eq:w.error.square.GRZALMS}, the above optimization problem becomes:
{
\begin{equation}\label{eq:minimiGRZA}
\begin{split}
	\{\mu^\star_n, \rho^\star_n\}&\!=\!
    \arg\min_{\mu, \rho} \, \tr\{\bQ_{n+1}\} \\
	&\!=\!\arg\min_{\mu, \rho} \,
    \tr\{\bQ_n\}\!+\!\mu^2g\!+\!\rho^2h\!
    +\!2\mu\rho \ell\!-\!2\mu r_1\!-\!2\rho r_2.
\end{split}
\end{equation}}
Equivalently, equation \eqref{eq:minimiGRZA} can be written in matrix form as:
\begin{equation}
              \label{eq:xi}
              \xi_{n+1} =[\mu\,\rho]\,\bH\,[\mu\;\rho]^\top
              - 2\,[r_1\; r_2]\,[\mu\;\rho]^\top + \xi_n,
 \end{equation}
which is a quadratic function of ${[\mu\; \rho]}$, with $\bH=\left[\begin{array}{cc} g & \ell \\\ell &h \end{array}\right]$.


By decomposing $g$ with respect to the two additive terms in the right-hand side of \eqref{eq:GRZA.a}, one can show that the Hessian matrix $\bH$ can be written as the sum of a covariance matrix and a positive semidefinite matrix. Matrix $\bH$ is thus positive semidefinite. In practice, since a covariance matrix is almost always positive definite~\cite{haykin2005}, we shall assume that $\bH$ is positive definite, which allows us to obtain the optimal parameters via:
\begin{equation}
          [\mu^\star_n\; \rho^\star_n]^\top =  \bH^{-1} [r_1\; r_2]^\top.
\end{equation}
Some elementary algebra leads to:
\begin{align}
	\mu^\star_n &= \frac{hr_1-\ell r_2}
    {gh-\ell^2}\label{eq:update.mu.ZALMS}\\
	\rho^\star_n &=\frac{gr_2-\ell r_1}
    {gh-\ell^2}\label{eq:update.rho.ZALMS}.
\end{align}
\noindent This result cannot be used in practice since it requires statistics that are not available in online learning scenarios. We shall now introduce an approximation for these quantities. Time index~$n$ is added in variables $g_n$, $h_n$, $\ell_n$, ${{r_1}_n}$ and ${{r_2}_n}$ for clearance.

Consider first the quantity $g_n$. With A1 and A2-A2', we obtain:
\begin{align}
        g_n & = \sigma_z^2\,\tr\{\sigma_u^2\bI\} + \tr\Bigl\{2\bR_u\bQ_n\bR_u + \tr\{\bR_u\bQ_n\}\bR_u\Bigr\}\notag\\
        & = \sigma_z^2 \sigma_u^2\,L + (2 + L)\sigma_u^2\,\zeta_n.  \label{eq:a.approximation}
\end{align}
\noindent Next, using A1 with $r_{1,n}$ yields:
\begin{align}
        r_{1,n} & = \zeta_n.\label{eq:p1.approximation}
\end{align}
\noindent Then, approximating the expectations in~\eqref{eq:GRZA.b}, \eqref{eq:GRZA.c} and \eqref{eq:GRZA.p2} by their instantaneous argument yields:
\begin{align}
	h_n & \approx (\bbeta_{n}\circ\bs_{n})^\top(\bbeta_{n}\circ\bs_{n})	\label{eq:b.Approximation}\\
	\ell_n & \approx \btw_n^\top\bu_n{\bu_{n}^\top(\bbeta_{n}\circ\bs_{n})} 	\label{eq:c.Approximation}\\
	r_{2,n} & \approx {(\bbeta_n\circ\bs_n)}^\top\tilde\bw_{n}.	\label{eq:p2.Approximation}
 \end{align}
{Now we construct an approximation for $\bw^\star$ at time instant $n$ in order to evaluating the weight error vector $\tilde\bw_n$. As proposed in~\cite{chen2015diffusion}, one strategy is to use a one-step approximation of the form:
\begin{equation}
	\hat{\bw}^\star_n
    = \bw_n - \eta_n\nabla J (\bw_n)
\end{equation}
with $\eta_n$ a positive step size to be determined. Given $\xi_n$, we seek~$\eta_n$ that minimizes $\xi_{n+1}$. Following the same reasoning as~\eqref{eq:emse}--\eqref{eq:update.rho.ZALMS} leads to $\eta_n=r_{1,n}/g_n$. Then, we approximate the gradient $\nabla J (\bw_n)$ with the instantaneous value $-e_{n}\bu_{n}$. Finally, we obtain the one-step approximation $\hat{\bw}^\star_n = \bw_n - \bp_{n}$ with $\bp_{n}= -\frac{r_{1,n}}{g_n}e_{n}\bu_{n}$.}

Quantities $g_n$ and $r_{1,n}$ depend on the EMSE $\zeta_n$, which is not available. Given~\eqref{eq:mse.ZALMS}, we suggest to use the following estimator:
\begin{align}
	\hat{\zeta}_n & = \max\{\hat e_n^2 - \sigma_v^2,\, 0\},\label{eq:zeta} \\
	\text{where:}\quad
    \hat e_n & = (1-\gamma)e_n + \gamma\hat e_{n-1}\label{eq:epsilon}
\end{align}
which provides an instantaneous approximation of the EMSE, with $\gamma$ a temporal smoothing factor in the interval $[0,1)$. To further improve the estimation accuracy of $\zeta_n$, we use $\zeta_{n_{\min}} = \sigma_u^2\,\tr\{\bQ_{n}\}$ calculated with~\eqref{eq:minimiGRZA} as a lower bound for $\zeta_n$. Indeed, since we minimized $\tr\{\bQ_{n}\}$ with respect to $\{\mu,\rho\}$ at iteration $n-1$, $\hat{\zeta}_n$ cannot be less than $\zeta_{n_{\min}}$ due to the approximation introduced in the derivation and the inherent variability of signal realizations. This, rather than \eqref{eq:zeta}, we suggest to use:
\begin{equation}
	\hat\zeta_n = \max\bigl\{\hat e_n^2 - \sigma_v^2,\,
    \zeta_{n_{\min}}\bigr\}.\label{eq:zetaNew}
\end{equation}
Note that non-negativity of $\mu$ and $\rho$ is also required. We did not consider this constraint in \eqref{eq:minimiGRZA} in order to get closed-form solutions as given by \eqref{eq:update.mu.ZALMS} and \eqref{eq:update.rho.ZALMS}.
We further impose a temporal smoothing over parameters $\mu_n^\star$ and $\rho_n^\star$, as well as a possible upper bound $\mu_{\max}$ for the step size $\mu$ to ensure the stability of the algorithm:
\begin{align}
	\mu_n &= \min\left\{\gamma'\mu_{n-1}
    + (1-\gamma')\mu_n^\star,\,\mu_{\max}\right\}
    \label{eq:mu.smoothing} \\
	\rho_n &= \gamma'\rho_{n-1}
    + (1-\gamma')\rho_n^\star\label{eq:rho.smoothing}
\end{align}
with $\gamma'$ a temporal smoothing factor in $[0,1)$.

\section{SIMULATION RESULTS}
\label{sec:simulation}

We shall now present simulation results to illustrate the effectiveness of our algorithms in non-stationary system identification scenarios. In all the experiments, the initial weight vector $\bw_0$ was set to the all-zero vector $\cb0_L$. The MSD learning curves were obtained by averaging results over 100 Monte-Carlo runs. The VP-GZA-LMS and VP-GRZA-LMS were compared with the standard LMS, GZA-LMS, GRZA-LMS, ZA-VSSLMS~\cite{Salman2012}, and WZA-VSSLMS~\cite{Salman2012} algorithms. Note that the last two algorithms are variable step size algorithms for general sparse system identification. For group-sparse algorithms, the group size $|\cp{G}_j|$ was set to 5 for all $j$, and $\varepsilon$ in \eqref{eq:twocase} was set to $0.1$. We set the parameters of all the algorithms so that the initial convergence rate of their MSD was almost the same. Two experiments were performed to compare their tracking behavior and steady-state performance with uncorrelated and correlated inputs.

In the first experiment, we considered a zero-mean white Gaussian input signal $u_n$ to be consistent with A2-A2'. The variance of $u_n$ was set to $\sigma_u^2=1$. The additive noise $z_n$ was an i.i.d.\! zero-mean white Gaussian noise with variance $\sigma_z^2=0.01$. The order of the unknown time-varying system was set to $L=35$. At time instant $n=1$, $8000$ and $16000$, we set the system parameter vector to $\bw^\star_1$, $\bw^\star_2$ and $\bw^\star_3$, respectively. Parameter vector $\bw^\star_2$ was a non-sparse vector, while $\bw^\star_1$  and $\bw^\star_3$ were group-sparse. They were defined as:
\begin{align*}
    \bw^\star_1 &= [0.8, \,0.5, \,0.3, \,0.2, \,0.1, \,\cb{0}_{15}, \,-0.05, \,-0.1, \,-0.2, \,-0.3,  -0.5,\, \cb{0}_{5},  \,0.5, \,0.25, \,0.5, \,-0.25, \,-0.5]^{\top};\notag
    \vspace{-3mm}
\end{align*}
\begin{align*}
        \bw^\star_2 &= [0.9, \,0.8, \,0.7, \,0.6, \,0.5, \,0.4, \,0.3, \,0.2, \,0.1, \,\cb{1}_{17}, \,-0.1, \,-0.2, -0.3, \,-0.4, \,-0.5, \,-0.6, \,-0.7, \,-0.8 \,-0.9]^{\top};\notag
        \vspace{-3mm}
\end{align*}
\begin{align*}
        \bw^\star_3 &= [1.2, \,0.9, \,0.8, \,0.7, \,0.6, \,0.5, \,0.4, \,0.2, \,0.5, \,0.4, \,\cb{0}_{15}, \,-0.4, -0.5, \,-0.2, \,-0.4, \,-0.5, \,-0.6 \,-0.7, \,-0.8, \,-0.9, \,-1.2]^{\top}.
\end{align*}

\begin{figure}[htb]
\begin{minipage}[b]{1.0\linewidth}
  \centering
  \centerline{\includegraphics[width=7.9cm]{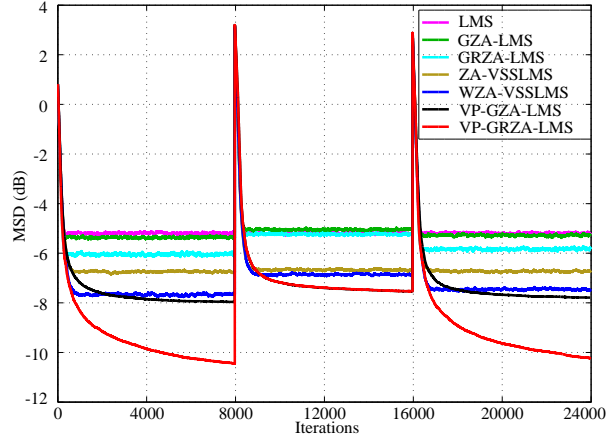}}
\end{minipage}
\caption{MSD learning curves for experiment 1 (white input).}
\label{fig1}
\end{figure}

\noindent The results are provided in Fig.~\ref{fig1}. Observe that all the algorithms outperformed the LMS in stages $\bw^\star_1$ and $\bw^\star_3$, demonstrating their effectiveness for group-sparse system identification. VP-GZA-LMS and VP-GRZA-LMS algorithms converged as fast as the other algorithms when estimating $\bw^\star_1$ but reached a lower misadjustement error, especially for VP-GRZA-LMS. The estimation of the non-sparse $\bw^\star_2$ caused a moderate performance degradation, mainly in their convergence speed. Indeed, their convergence speeds slowed down compared to the other algorithms but they however reached a smaller MSD. The estimation of $\bw^\star_3$ confirms improved performance and tracking ability of VP-GZA-LMS and VP-GRZA-LMS.

\begin{figure}[htb]
\begin{minipage}[b]{1.0\linewidth}
  \centering
  \centerline{\includegraphics[width=7.9cm]{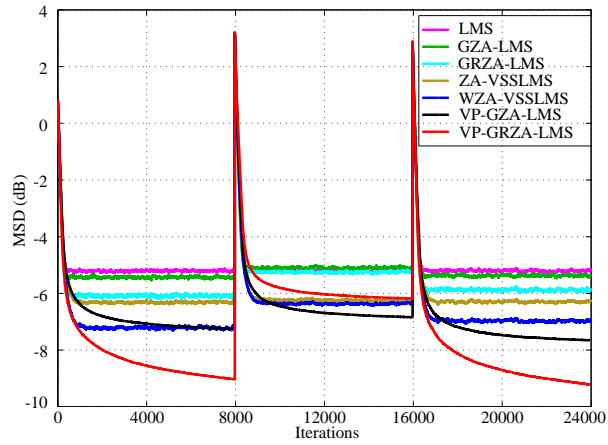}}
\end{minipage}
\caption{MSD learning curves for experiment 2 (non-Gaussian colored input).}
\label{fig2}
\end{figure}

\begin{figure}[htb]
\begin{minipage}[b]{1.0\linewidth}
  \centering
  \centerline{\includegraphics[width=7.9cm]{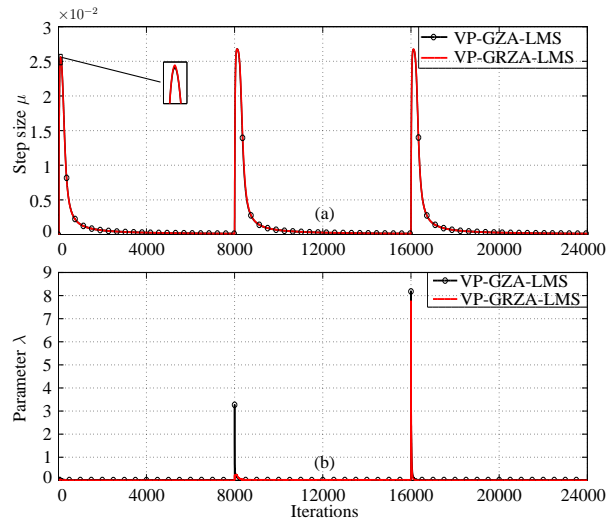}}
\end{minipage}
\caption{(a) Evolution of the step size $\mu$ and (b) the regularization parameter $\lambda$ of VP-GZA-LMS and VP-GRZA-LMS algorithms for experiment 2.}
\label{fig3}
\end{figure}

In the second experiment, we used the same setting except that the input signal was generated with a first-order AR process given by $u_n = \alpha\,u_{n-1} + v_n$, where $v_n$ was an i.i.d.\! zero-mean random variable distributed according to the following Gaussian mixture distribution: $0.5\,\cp{N}(a\cdot\sigma_v,\sigma_v^2)+0.5\,\cp{N}(-a\cdot\sigma_v,\sigma_v^2)$. Its parameters were set to $a=3/2$ and $\sigma_v^2=4/13$, so that $\sigma^2_u=1$. The correlation coefficient of $u_n$ was set to $\alpha=1/2$. In this way, assumptions A2--A2' were both relaxed in order to test the robustness of our approach. The learning curves of all the algorithms are provided in Fig.~\ref{fig2}. The evolution of the step size and the regularization parameter over time of VP-GZA-LMS and VP-GRZA-LMS are provided in Fig.~\ref{fig3}. Though there was some performance degradation of VP-GZA-LMS and VP-GRZA-LMS algorithms compared with the first experiment, the VP-GRZA-LMS algorithm still led to the lowest steady-state MSD along with the fastest convergence speed among all the competing algorithms for $\bw^\star_1$ and $\bw^\star_3$. The performance of VP-GZA-LMS was almost at the same level as the best of the competing algorithms. Despite the loss of assumptions A2--A2', VP-GZA-LMS and VP-GRZA-LMS algorithms still worked well with non-Gaussian correlated inputs. The results in Fig.~\ref{fig3} show that VP-GZA-LMS and VP-GRZA-LMS set the step size and regularization parameter to large values at the beginning of each estimation stage in order to ensure tracking ability and promote sparsity. Then they gradually reduced them to ensure small MSD.

\section{CONCLUSION}
\label{sec:conclusions}
In this paper, we introduced VP-GZA-LMS and VP-GRZA-LMS algorithms to address online group-sparse system identification problems. Based on a model of the transient behavior of the GRZA-LMS algorithm, we proposed to minimize the MSD with respect to the step size and regularization parameter, simultaneously, at each iteration. This led to a convex optimization problem with a closed-form solution. Simulation results demonstrated the effectiveness of VP-GZA-LMS and VP-GRZA-LMS algorithms over other existing variable step size methods, even for non-Gaussian correlated inputs.

\balance
\newpage

\bibliographystyle{IEEE}
\bibliography{ref}

\begin{thebibliography}{10}

\bibitem{Sayed2008}
A.~H. Sayed,
\newblock {\em Adaptive Filters},
\newblock John Wiley \& Sons, Inc., 2008.

\bibitem{Widrow1985}
B.~Widrow and S.~D. Stearns,
\newblock {\em Adaptive Signal Processing},
\newblock Prentice-Hall, Inc., Upper Saddle River, NJ, USA, 1985.

\bibitem{Duttweiler2000}
D.~L. Duttweiler,
\newblock ``Proportionate normalized least-mean-squares adaptation in echo
  cancelers,''
\newblock {\em IEEE Transactions on Speech and Audio Processing}, vol. 8, no.
  5, pp. 508--518, Sep. 2000.

\bibitem{Benesty2002}
J.~Benesty and S.~L. Gay,
\newblock ``An improved {PNLMS} algorithm,''
\newblock in {\em Proceedings of the IEEE ICASSP}, May 2002, vol.~2, pp.
  II--1881--II--1884.

\bibitem{Chen2009ZALMS}
Y.~Chen, Y.~Gu, and A.~O. Hero,
\newblock ``Sparse {LMS} for system identification,''
\newblock in {\em Proceedings of the IEEE ICASSP}, Taipei, China, Apr. 2009,
  pp. 3125--3128.

\bibitem{Chen2012Recursive}
Y.~Chen and A.~O. Hero,
\newblock ``Recursive $\ell_{1,\infty}$ group lasso,''
\newblock {\em IEEE Transactions on Signal Processing}, vol. 60, no. 8, pp.
  3978--3987, Aug. 2012.

\bibitem{Chen2010RegularizedLMS}
Y.~Chen, Y.~Gu, and A.~O. Hero,
\newblock ``Regularized least-mean-square algorithms,''
\newblock 2010, Arxiv preprint arXiv:1012.5066.

\bibitem{Schreiber1995}
W.~F. Schreiber,
\newblock ``Advanced television systems for terrestrial broadcasting: Some
  problems and some proposed solutions,''
\newblock {\em Proceedings of the IEEE}, vol. 83, no. 6, pp. 958--981, Jun.
  1995.

\bibitem{Huang2010}
J.~Huang and T.~Zhang,
\newblock ``The benefit of group sparsity,''
\newblock {\em The Annals of Statistics}, vol. 38, no. 4, pp. 1978--2004, Aug.
  2010.

\bibitem{Paleologu2008}
C.~Paleologu, J.~Benesty, and S.~Ciochin{\u{a}},
\newblock ``A variable step-size proportionate { NLMS } algorithm for echo
  cancellation,''
\newblock {\em Revue Roumaine des Sciences Techniques -- Serie Electrotechnique
  et Energetique}, vol. 53, pp. 309--317, 2008.

\bibitem{Salman2012}
M.~S. Salman, M.~N.~S. Jahromi, A.~Hocanin, and O.~Kukrer,
\newblock ``A zero-attracting variable step-size {LMS} algorithm for sparse
  system identification,''
\newblock in {\em IX International Symposium on Telecommunications (BIHTEL)},
  Oct. 2012, pp. 1--4.

\bibitem{Fan2017}
T.~Fan and Y.~Lin,
\newblock ``A variable step-size strategy based on error function for sparse
  system identification,''
\newblock {\em Circuits, Systems, and Signal Processing}, vol. 36, no. 3, pp.
  1301--1310, Mar. 2017.

\bibitem{Benesty2006}
J.~Benesty, H.~Rey, L.~R. Vega, and S.~Tressens,
\newblock ``A nonparametric {VSS NLMS} algorithm,''
\newblock {\em IEEE Signal Processing Letters}, vol. 13, no. 10, pp. 581--584,
  Oct 2006.

\bibitem{Kwong1992}
R.~H. Kwong and E.~W. Johnston,
\newblock ``A variable step size {LMS} algorithm,''
\newblock {\em IEEE Transactions on Signal Processing}, vol. 40, no. 7, pp.
  1633--1642, Jul. 1992.

\bibitem{Chen2016ZALMS}
J.~Chen, C.~Richard, Y.~Song, and D.~Brie,
\newblock ``Transient performance analysis of zero-attracting {LMS},''
\newblock {\em IEEE Signal Processing Letters}, vol. 23, no. 12, pp.
  1786--1790, Dec. 2016.

\bibitem{haykin2005}
S.~Haykin,
\newblock {\em Adaptive Filter Theory},
\newblock Pearson Education India, 4th edition, 2005.

\bibitem{chen2015diffusion}
J.~Chen, C.~Richard, and A.~H. Sayed,
\newblock ``Diffusion {LMS} over multitask networks,''
\newblock {\em IEEE Transactions on Signal Processing}, vol. 63, no. 11, pp.
  2733--2748, 2015.

\end{thebibliography}

\balance
\end{document}